\documentclass[runningheads]{llncs}
\usepackage[T1]{fontenc}
\usepackage{graphicx}
\usepackage{mdframed}
	\usepackage{listings} 
	\usepackage{amsmath}
    \usepackage{subcaption}
    \usepackage{forest}
    \usepackage{booktabs}
    \usepackage[dvipsnames]{xcolor}
    \usepackage{framed}

\usepackage{changepage}
\newcommand{\interviewquote}[1]{
 \def\FrameCommand{
    \hspace{0pt} 
    {\color{cyan}  \vrule width 2pt}%
    {\color{white} \vrule width 3pt}%
  }%
  \MakeFramed{\advance\hsize-\width\FrameRestore}%
  \noindent
  \begin{adjustwidth}{}{1pt}
  {\footnotesize``#1''}
  \end{adjustwidth}
  \endMakeFramed%
}
\begin{document}
\title{Supporting Industrial Test-Failure Analysis with LLM-Based Systems: An Experience Report}
\titlerunning{Supporting Industrial Test-Failure Analysis with LLM-Based Systems}
%
\author{Eric Jansson\inst{1} \and
Per Strandberg\inst{2}\orcidID{0000-0003-1688-6937} \and
Thomas Sörensen\inst{2} \and
Eduard Paul Enoiu\inst{1}\orcidID{0000-0003-2416-4205} \and
Wasif Afzal\inst{1}\orcidID{0000-0003-0611-2655}}
\authorrunning{E.\ Jansson et al.}
%
\institute{Mälardalen University, Västerås, Sweden 
\email{eric95.jansson@gmail.com, \{eduard.paul.enoiu,wasif.afzal\}@mdu.se}
\and
Westermo Network Technologies AB, Västerås, Sweden
\email{\{per.strandberg,thomas.sorensen\}@westermo.com}\\
\url{https://www.westermo.com}
}
\maketitle              
%
%
%
%
%
%
%
%
%
%
%
%


\begin{abstract}

This study examines tool-augmented Large Language Model (LLM) systems for supporting Root Cause Analysis (RCA) of nightly test failures at Westermo Network Technologies AB. Nightly test executions produce heterogeneous test data and logs that practitioners currently inspect manually across multiple sources.
We implemented an RCA workflow in single-agent and orchestrated multi-agent configurations, both with access to test metadata and logs. An exploratory industrial case study used two real failure scenarios. Six practitioners evaluated the scenario reports through a survey and focus group, and operational measurements were collected from 120 repeated executions. The evaluation covered practitioner-perceived correctness, reasoning quality, fix realism, clarity, usefulness, and trust, as well as cost, duration, and consistency. Neither configuration showed a consistent practitioner-perceived quality advantage across the two scenarios. The single agent system generated reports faster and at lower cost, making it the more practical baseline in this context. The potential benefits of agent architectures require further evaluation in more complex scenarios. 

\end{abstract}

\section{Introduction}
\label{sec:intro}

Logs are used to understand system behavior and diagnose failures \cite{xue2024review}. As systems scale, log volume and diversity increase, making analysis harder. In this paper, multi-source test data denotes test metadata and log evidence
retrieved from the controller and multiple devices under test. The sources may differ in schema, message format and content.

Automated log analysis can reduce manual effort but often relies on fixed patterns or log formats, limiting its use with varied or domain-specific logs \cite{xue2024review}. LLMs can process both textual and semi-structured data, including system logs \cite{LLM_a_review}. For complex tasks requiring information retrieval and multi-step analysis, they can also be combined with external tools and structured reasoning \cite{roy2024exploringAgents}.

Root cause analysis (RCA) of test failures aims to determine why a test failed and what should be investigated next. In industrial testing, this often requires engineers to examine detailed execution logs and connect events across the test environment~\cite{hermawan2025benchmarking}. RCA involves detecting abnormal events and interpreting these in relation to the test context and expected system behavior. Because failures may originate from misconfiguration, timeouts, aborted tests, and interactions between system components, several sources of evidence may be required. Logs can show what happened, while monitoring data, topology information, and expert knowledge may be needed to explain why it happened~\cite{zhang2021cloudrca}.

LLM-based agents can retrieve external data and use it in iterative test log analysis, where evidence may be distributed across several devices \cite{agent_tools_architectures}. A single agent handles the full task, while a multi-agent system divides it among specialized agents. Even if this may improve separation of responsibilities, it also adds coordination overhead that could affect report clarity, usefulness, trust, cost and execution time. Evidence in using both configurations on the same industrial RCA task remains limited.

\begin{mdframed}[linewidth=1pt, linecolor=black, backgroundcolor=white!10, roundcorner=5pt]
\textbf{Industrial Context:} At Westermo Network Technologies AB (hereafter, Westermo), nightly regression tests exercise network devices through an internal test framework. When a test fails, practitioners inspect test metadata, identify the devices involved and search controller and device logs to determine if the failure originates in product software, the test framework, the test environment or their interaction. Relevant evidence may be distributed across several devices and time windows, and the current workflow therefore relies heavily on practitioner experience.
\end{mdframed}

This paper reports industrial experience that examines which of the two concrete configurations provides the more appropriate baseline for the current task and what changes would be needed before integration into routine test-failure analysis at Westermo. Once given a failure case, both configurations autonomously retrieve and analyze test data and logs and generate an RCA report for practitioner assessment and further investigation. Using an exploratory industrial case study and mixed-methods design, the architectures were evaluated through system measurements, scenario-based surveys and a focus group. The evaluation considered cost, duration, consistency and practitioners' perceptions of the usefulness and trustworthiness of the generated RCA reports. 

The paper contributes: (1) a tool-augmented RCA workflow instantiated in single-agent and orchestrated multi-agent configurations; (2) an industrial evaluation of these configurations using the same model, data sources, tools and report format based on two real failure scenarios, feedback from six practitioners and 120 repeated executions; and (3) practitioner-grounded lessons concerning evidence presentation, workflow integration, uncertainty and the operational trade-offs of agent specialization.
This paper is based upon the Master's thesis work of the first author
\cite{jansson2026aidriven}.

\section{Background}
\label{sec:background}

Automated log analysis often transforms unstructured messages into structured event templates. However, template-based methods can be sensitive to changing formats, heterogeneous components, and minor wording differences, limiting their use in dynamic industrial environments~\cite{xue2024review}.

LLMs can process natural language and semi-structured data, including system logs, making them suitable for test-failure analysis~\cite{LLM_a_review}. However, effective analysis requires domain-specific context that may not be available in the model's training data, such as test metadata, device mappings, logs, topology information, and technical documentation. This information can be supplied through external tools or retrieval mechanisms~\cite{RAG_vs_finetuning}. Complex test-log analysis may therefore require several interactions in which the system retrieves evidence, interprets the available context, and generates a structured RCA report~\cite{roy2024exploringAgents}.

\section{Related Work}

Prior log-analysis methods have used structured classification, semantic representations, and fine-tuned language models for fault diagnosis and anomaly detection~\cite{almodovar2024logfit,zhang2019robust,zou2016uilog}. Other approaches combine multiple evidence sources, such as logs, performance indicators, and topology information, to support root cause analysis~\cite{zhang2021cloudrca}. LLM-based methods have also been applied to log interpretation and fault localization~\cite{qi2023loggpt,shan2024faceityourselves}. However, these studies mainly detect abnormal events or localize faults, whereas this study evaluates complete RCA reports intended to help practitioners assess evidence and continue an investigation.

Recent work has used LLMs and tool-augmented agents to support RCA more directly. RCACopilot collects diagnostic information before generating root-cause predictions and explanations~\cite{chen2024automatic}, while ReAct-style agents and RCAgent retrieve evidence dynamically from external diagnostic services~\cite{roy2024exploringAgents,wang2024rcagent}. Structured and multi-agent approaches, including Flow-of-Action and MA-RCA, divide diagnosis into controlled steps or specialized responsibilities~\cite{fu2026ma-rca_leveraging,pei2025flow-of-action}. Most of these systems are evaluated in cloud or operational incident settings rather than industrial software testing, and they provide limited evidence about how practitioners assess the resulting reports.

Research outside RCA also suggests that architectural complexity is not automatically beneficial. SWE-agent shows that agent performance depends strongly on the design of tool interfaces and feedback~\cite{yang2024swe}, while AGENTLESS demonstrates that predefined workflows can remain competitive with more autonomous agent designs~\cite{xia2025demystifying}. These findings support treating architecture as a task-dependent design choice. Prior work provides limited comparisons between single-agent and multi-agent configurations on the same industrial RCA task, using the same data, tools, and report format. This study addresses that gap through practitioner evaluation and operational measurements of cost, duration, and semantic output stability.
Together, these works suggest that architecture should be treated as a design choice, where added specialization and context separation may help but also introduce coordination overhead and additional failure points. Overall, prior work demonstrates the value of tool use, workflow control and agent specialization for RCA. However, most studies concern cloud operations, microservices and incident triage. This motivates the examination of these approaches in industrial embedded system software testing, where the available evidence primarily consists of test metadata, device mappings, and logs.

\section{Study Design and Evaluation Process} 
\label{sec:method}

This method section provides an overview of the study design, including the system development process and an industrial case study at Westermo.


\begin{figure}[!ht]
  \centering
  \includegraphics[width=0.9\textwidth]{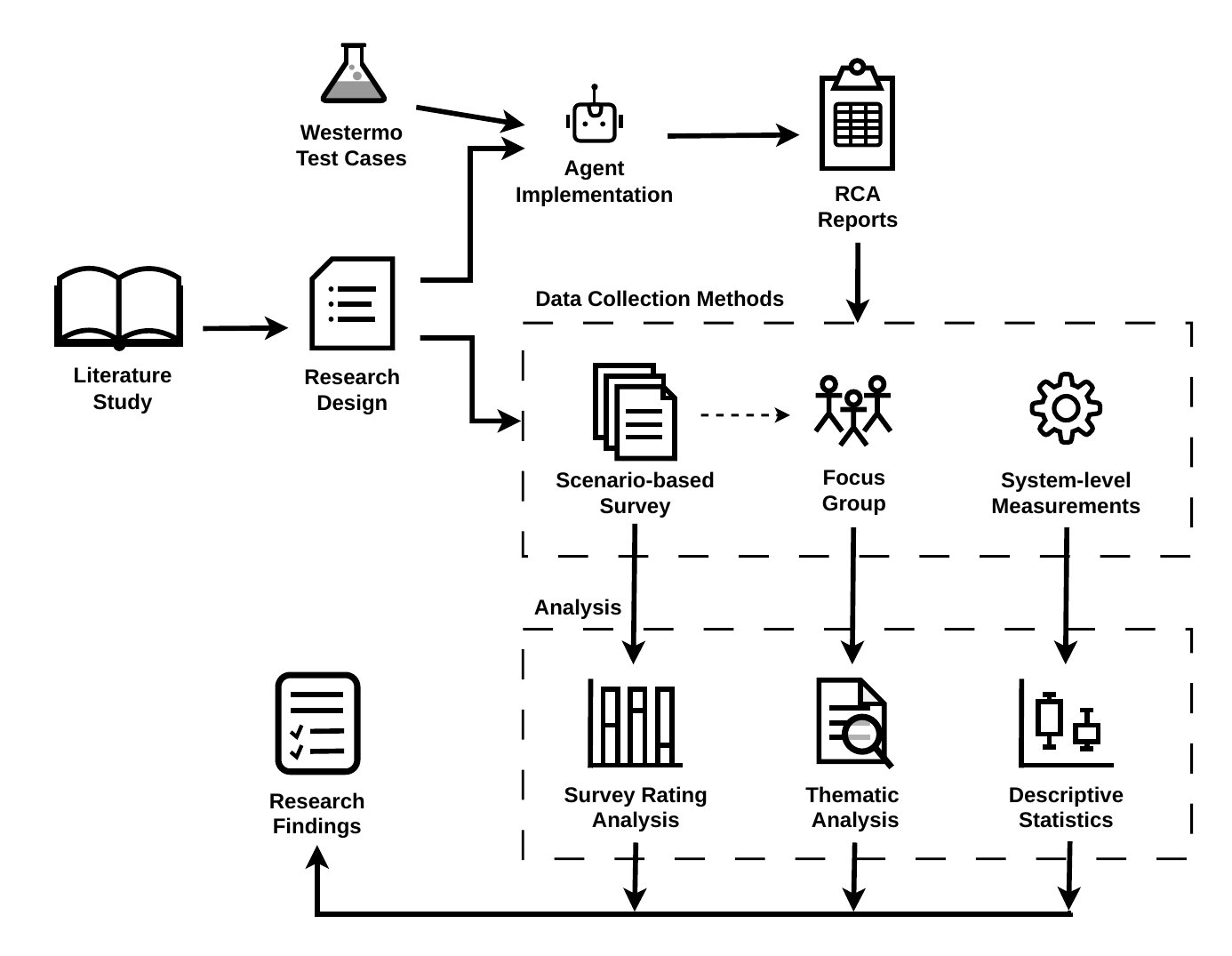}
  \caption{Overview of the study design and evaluation process.}
  \label{fig:thesis_process_overview}
\end{figure}

We conducted an exploratory industrial case study at Westermo using mixed-methods data collection~\cite{thematic_analysis_braun,runeson2009guidelines,mixed_methods_storey2025guiding}. As illustrated in Figure~\ref{fig:thesis_process_overview}, the literature review informed the system design, evaluation metrics and data collection methods. The implementation was constrained by Westermo's existing infrastructure, data formats and available test-failure scenarios. The evaluation dimensions were informed by previous work distinguishing correctness from explanation quality~\cite{anghel2025pearl}, using inference time as an efficiency measure~\cite{cui2024logeval} and relating useful explanations to failure reasoning and plausible corrective actions~\cite{kang2024quantitative}.

Four real test failures were selected in consultation with test framework experts at Westermo. Two failures were used during development to refine the prompts, tool instructions, and retrieval behavior. The remaining two were reserved for evaluation and used in the practitioner survey and focus group. For each evaluation scenario, both configurations received the same initial failure information and had access to the same external data sources. 

Six Westermo practitioners evaluated reports generated for the two evaluation scenarios through a survey followed by a semi-structured focus group. The participants represented test framework architecture and development, operating system development, release management, agile coordination and project management. The focus group lasted approximately 80 minutes and was conducted in a hybrid format, with five participants attending in person and one remotely. The survey assessed six dimensions: perceived correctness, reasoning quality, realism of the proposed fix, clarity, usefulness and trust. The focus group explored the reports' reasoning, the useful and misleading guidance, the level of detail, the structure, and the perceived differences between the single and multi-agent systems. With participant consent, the discussion was recorded and transcribed. The first author manually coded relevant segments using the predefined themes and identified recurring observations and differences between participants. Representative quotations were selected to explain patterns in the survey results.

System measurements were collected independently from the practitioner evaluation. Both architectures were executed 30 times for each of the two scenarios, producing 120 RCA reports in total. Three metrics were recorded. \emph{Cost} was calculated from Azure's \textit{CostByResource} data using input, cached-input and output-token usage obtained from execution traces. \emph{Duration} measured the time from initiating an analysis until the final report was produced. \emph{Consistency} represented the semantic stability of reports generated repeatedly from the same input. An LLM evaluator compared the \textit{Root Cause} and \textit{Evidence} sections and grouped reports with semantically similar conclusions. Consistency was calculated as the size of the largest similarity group, divided by the number of 30 reports generated for that scenario and architecture.


Survey ratings and system measurements were analyzed descriptively due to the small participant group and the exploratory study design. The qualitative findings were used to interpret the rating distributions and explain how practitioners assessed the reports in relation to their work.

\section{An Implementation of LLM-based Agents}
\label{sec:implementation}

This section describes two LLM-based systems for supporting RCA of nightly test failures at Westermo. The single-agent variant handled the entire workflow, while the multi-agent variant divided tasks between an orchestrator and specialized subagents. Both used the same data sources, tool categories, and report structure to enable a controlled comparison.

\begin{figure}[ht]
  \centering
  \includegraphics[width=0.95\textwidth]{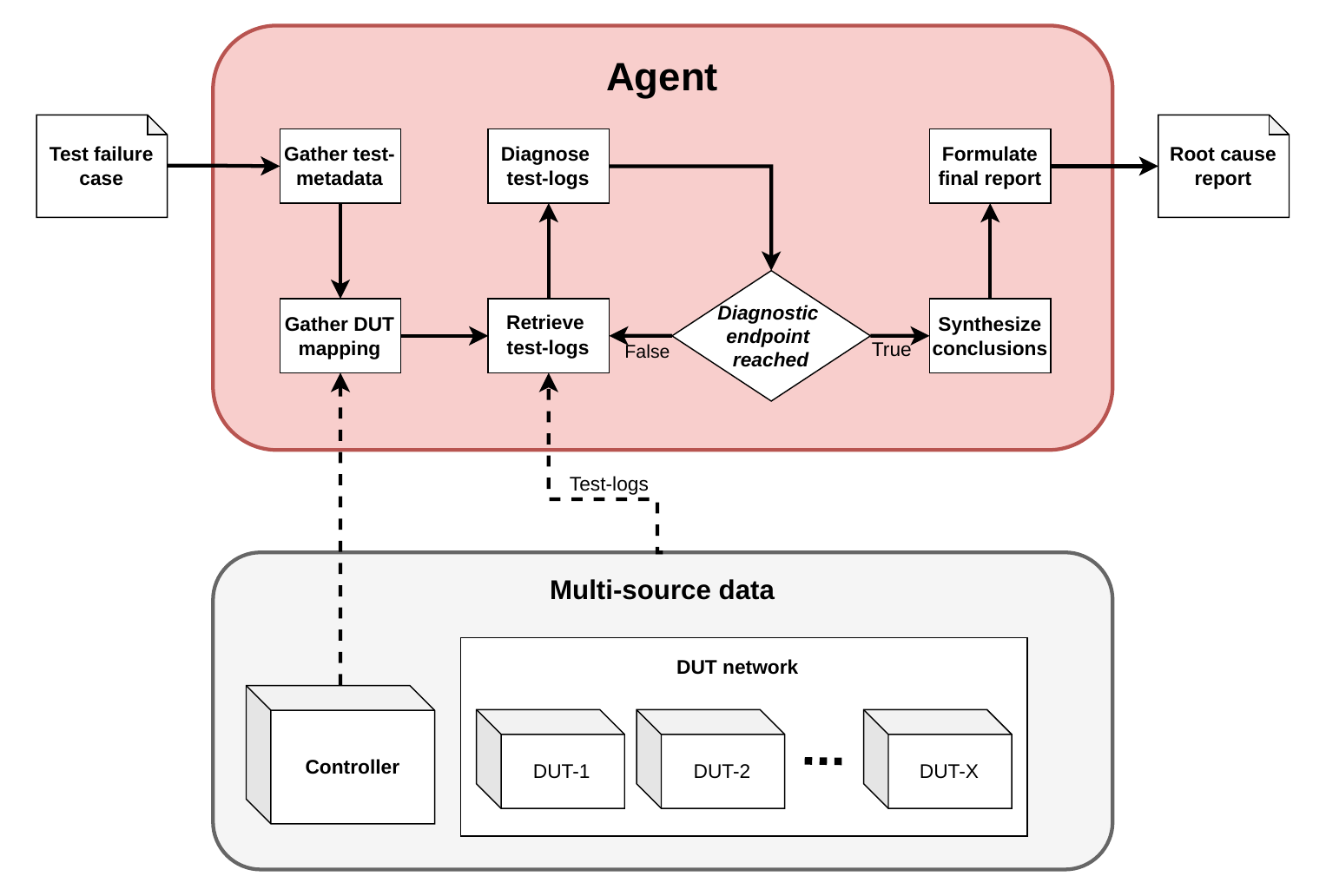}
  \caption{RCA workflow and multi-source test data used by the agent systems.}
  \label{fig:agent_workflow}
\end{figure}

\subsection{Agent Workflow}
\label{sec:agent_workflow}

Figure \ref{fig:agent_workflow} presents the workflow used by both agent systems. Starting from a failed test case, the agents retrieve test metadata, identify the relevant devices under test (DUTs), collect logs, analyze the evidence and generate an RCA report. Evidence may originate from the controller and several devices involved in the same failure.

External test data was accessed through two Model Context Protocol (MCP) servers: one for test result metadata and another one for test logs. Only tools needed for metadata, device mapping and log retrieval were exposed. In the single agent system, the main agent called these tools directly, while the multi-agent system assigned them to specialized subagents coordinated by an orchestrator. The internal MCP interfaces and retrieval parameters are described only at a high level because they are proprietary to Westermo.

The workflow was defined through prompts rather than enforced as a fixed sequence of code. This allowed the agents to select tools dynamically while following the same general process. After retrieving metadata and identifying the relevant DUTs, the agents iteratively collected and analyzed logs until they deemed the available evidence sufficient.

The final report followed a common structure. The \textit{Root Cause} field stated the most likely cause of the failure, while \textit{Confidence} indicated the agent's uncalibrated confidence in that conclusion. \textit{Evidence} presented the most relevant supporting log entries, and \textit{Reasoning Steps} explained how the evidence was interpreted. \textit{Next Steps} suggested actions for confirming, investigating, or resolving the failure. Finally, \textit{Assumptions} identified unverified conditions used in the analysis, while \textit{Limitations} described missing information or other constraints affecting the conclusion.

\subsection{Implementation Platform and Configuration}

Azure AI Foundry was selected because it aligned with Westermo's infrastructure and data protection requirements for processing internal logs, prompts, tool outputs and RCA reports. GPT-5.4 was selected because it was the model supported in Westermo's deployment and workflow at the time of the experiment.
%
The systems were implemented in Python using \textit{Deep Agents}\footnote{Deep Agents documentation: \url{https://docs.langchain.com/oss/python/deepagents}}, a LangChain-based framework supporting tool use, subagents, middleware and multi-step execution. Each system combined the Azure-hosted LLM, a restricted set of MCPs and custom tools, and prompts that define the RCA workflow.

%

\subsection{Prompt Design and Agent Responsibilities}
\label{sec:context_and_prompt_construction}

Both systems used modular prompts that included workflow instructions, tool descriptions, evidence-grounding rules, output requirements and domain knowledge of Westermo’s test logs. These modules were combined at runtime before the specific failure case was provided.

The prompt was initially developed for the single-agent system and then distributed across the multi-agent architecture. The \textit{orchestrator} received the overall workflow and coordination instructions, while each subagent received only the context needed for its responsibility. The \textit{metadata subagent} retrieved test information, the \textit{mapping subagent} identified the relevant DUTs and the \textit{log-analysis subagent} retrieved and interpreted log evidence. The \textit{RCA-report subagent} then combined the findings into the final structured report. Formatting instructions controlled how information was returned between the subagents and the orchestrator. The structure of prompts is discussed in section 6.6 of \cite{jansson2026aidriven}.

\subsection{Token Efficiency and Execution Monitoring}
\label{sec:token_usage_and_monitoring}

Reducing token usage was important because retrieved logs accounted for most of the model's context and directly affected costs. Log entries were therefore filtered to retain only timestamps and message content, reducing the log context by approximately 80\%. Query filters and retrieval limits further constrained repeated access to logs. \textit{Langfuse} was used to trace executions, inspect tool calls, debug unexpected behavior, and collect data on duration, cost, and token usage. During development, live monitoring was also used to identify repeated tool calls and cases where agents failed to progress toward a final report.

\section{Results}

This section reports findings from the survey, focus group and 120 system executions. Six experienced practitioners evaluated both scenarios, and each combination of scenario architectures was executed 30 times. 
While this section presents the summary of the results, further diagrams and details are available in the master's thesis report \cite{jansson2026aidriven}.

\subsection{Practitioner Survey Results}
\label{sec:results_survey}

The survey data is treated as one source of evidence within the mixed-methods evaluation. Given the small participant group and ordinal ratings, the results are presented descriptively using rating distributions and median differences. The symbol $\Delta$ denotes the difference between the multi-agent and single-agent median ratings.
\begin{figure}[htb]
  \centering
  \includegraphics[width=0.55\textwidth]{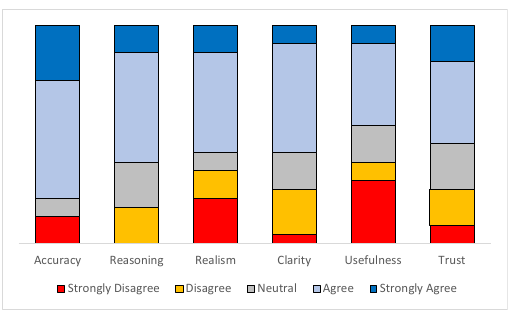}
  \caption{Overall survey rating distribution across all evaluated reports. Each metric includes 24 ratings from six participants evaluating four report combinations.}
  \label{fig:T12SMA}
\end{figure}
Figure \ref{fig:T12SMA} shows a slight tendency toward positive ratings, although strongly negative responses also occurred. All metrics except \textit{Reasoning} included both the lowest and highest ratings.
The results were scenario-dependent. The single-agent report received higher median ratings across all metrics in Scenario 1, while the multi-agent report received higher ratings in Scenario 2. The largest difference concerned \textit{Usefulness}, with $\Delta=-1.5$ in Scenario 1 and $\Delta=+2.0$ in Scenario 2. This crossover indicates no consistent practitioner-perceived advantage for either architecture. Although the single-agent report in Scenario 1 received more positive ratings, some negative responses were included. The single-agent report in Scenario 2 had the most negative overall rating distribution among the four report combinations.

\begin{figure}[h!t]
  \centering
  \begin{subfigure}[t]{0.48\textwidth}
    \centering
    \includegraphics[width=\linewidth]{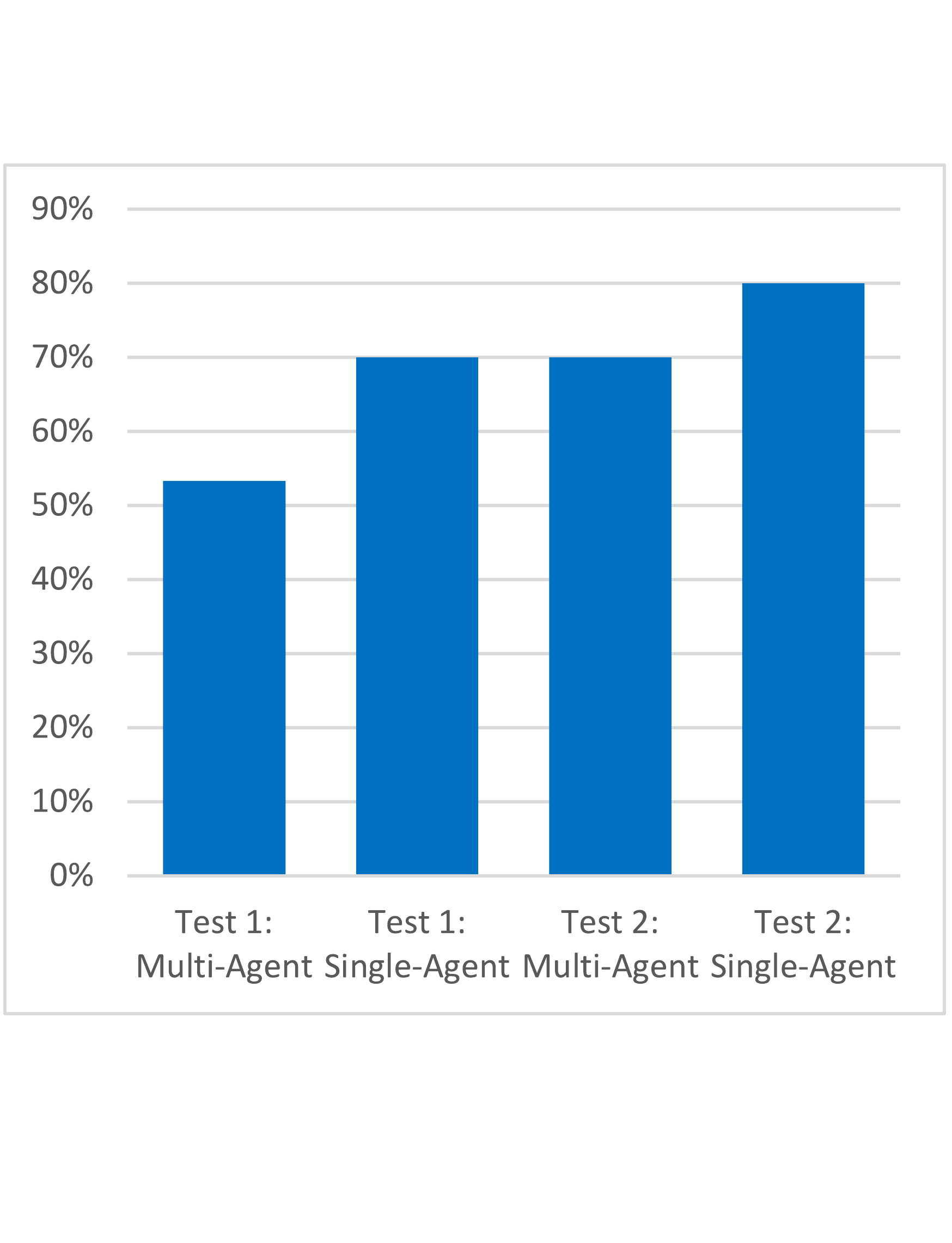}
    \caption{Largest consistency groups}
    \label{fig:consistency_results}
  \end{subfigure}
  \begin{subfigure}[t]{0.48\textwidth}
    \centering
    \includegraphics[width=\linewidth]{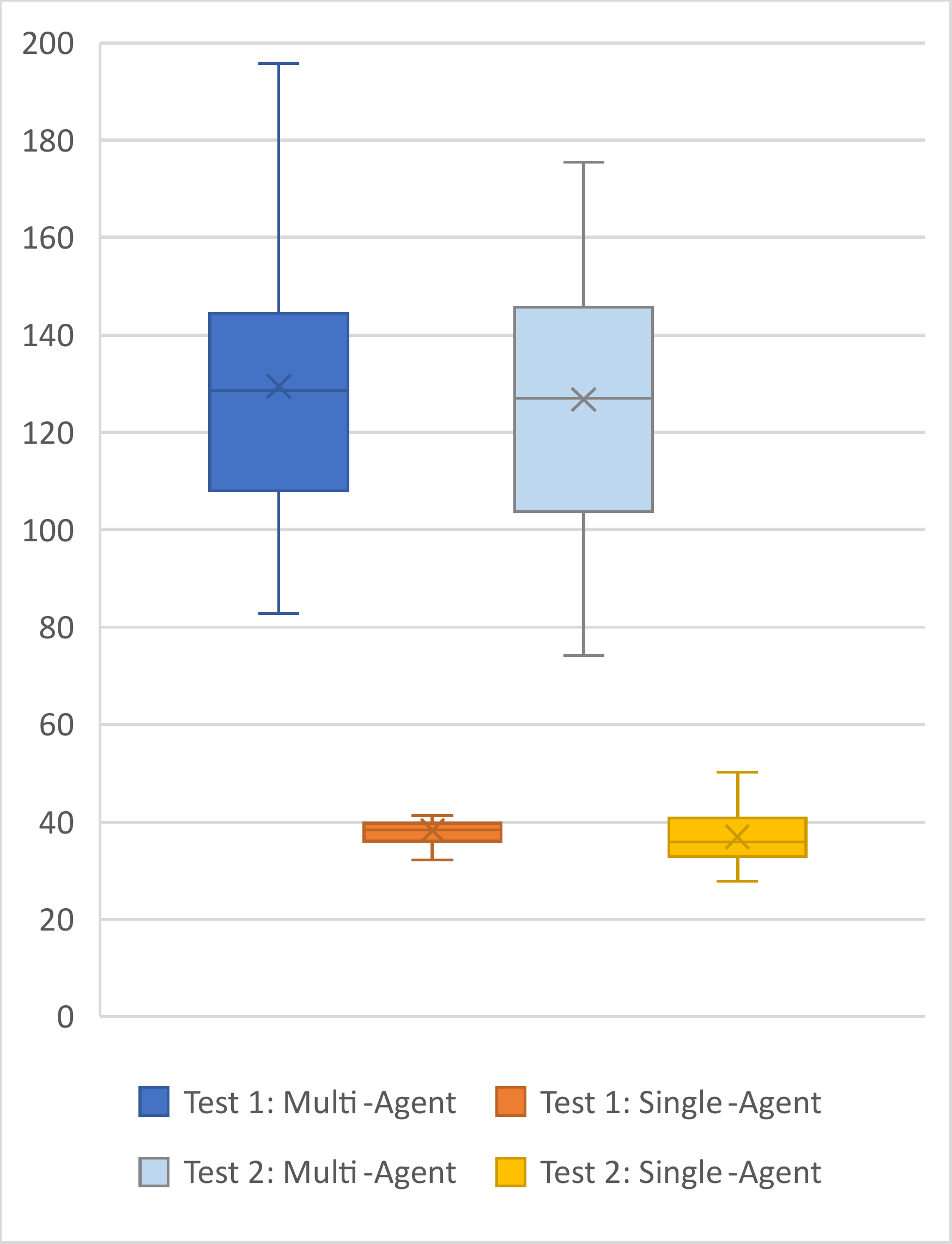}
    \caption{Duration (s)}
    \label{fig:duration}
  \end{subfigure}
    \begin{subfigure}{0.48\textwidth}
        \centering
        \includegraphics[width=\linewidth]{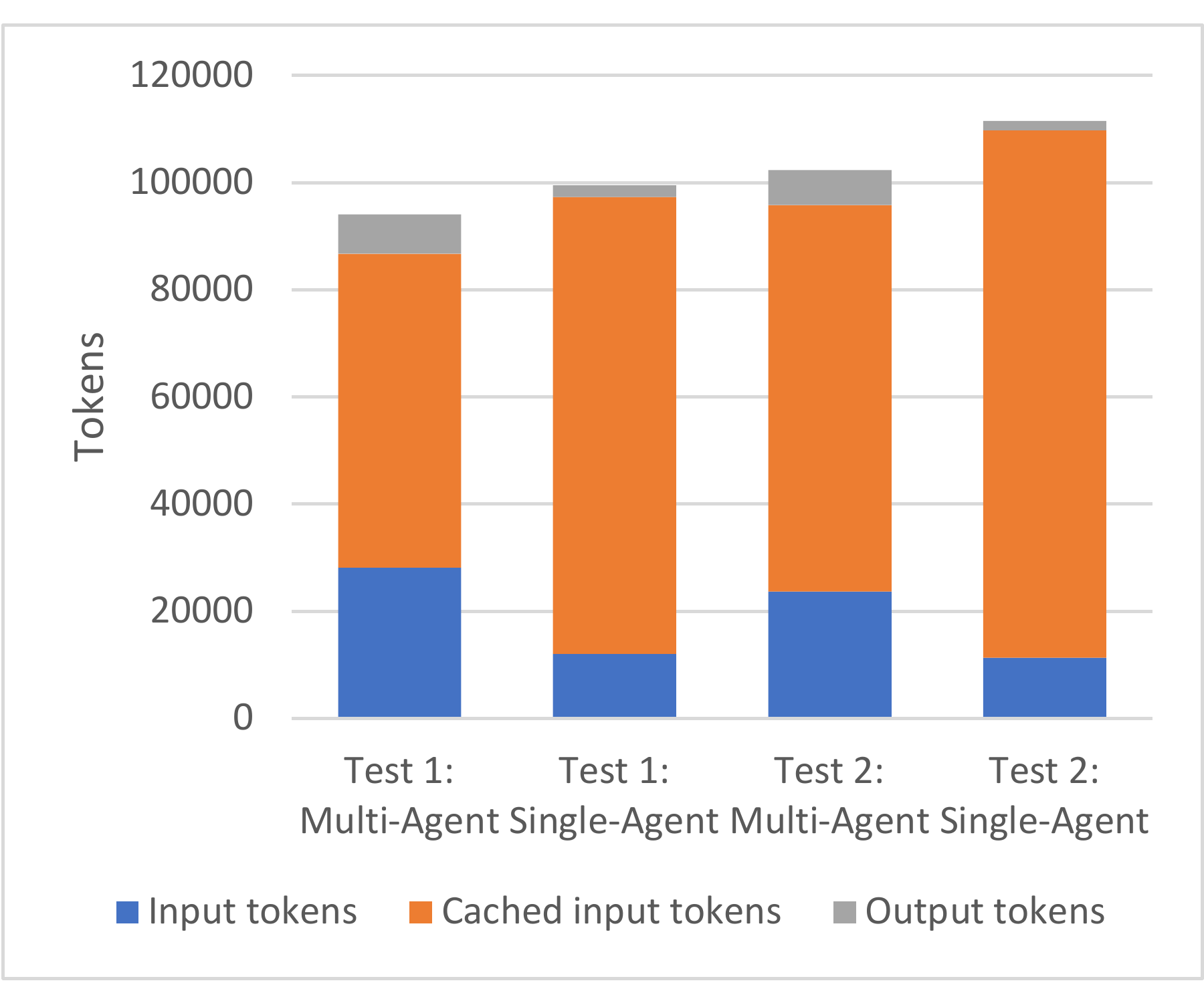}
        \caption{Token usage by token type}
        \label{fig:token_consumption}
    \end{subfigure}
    \begin{subfigure}{0.48\textwidth}
        \centering
        \includegraphics[width=\linewidth]{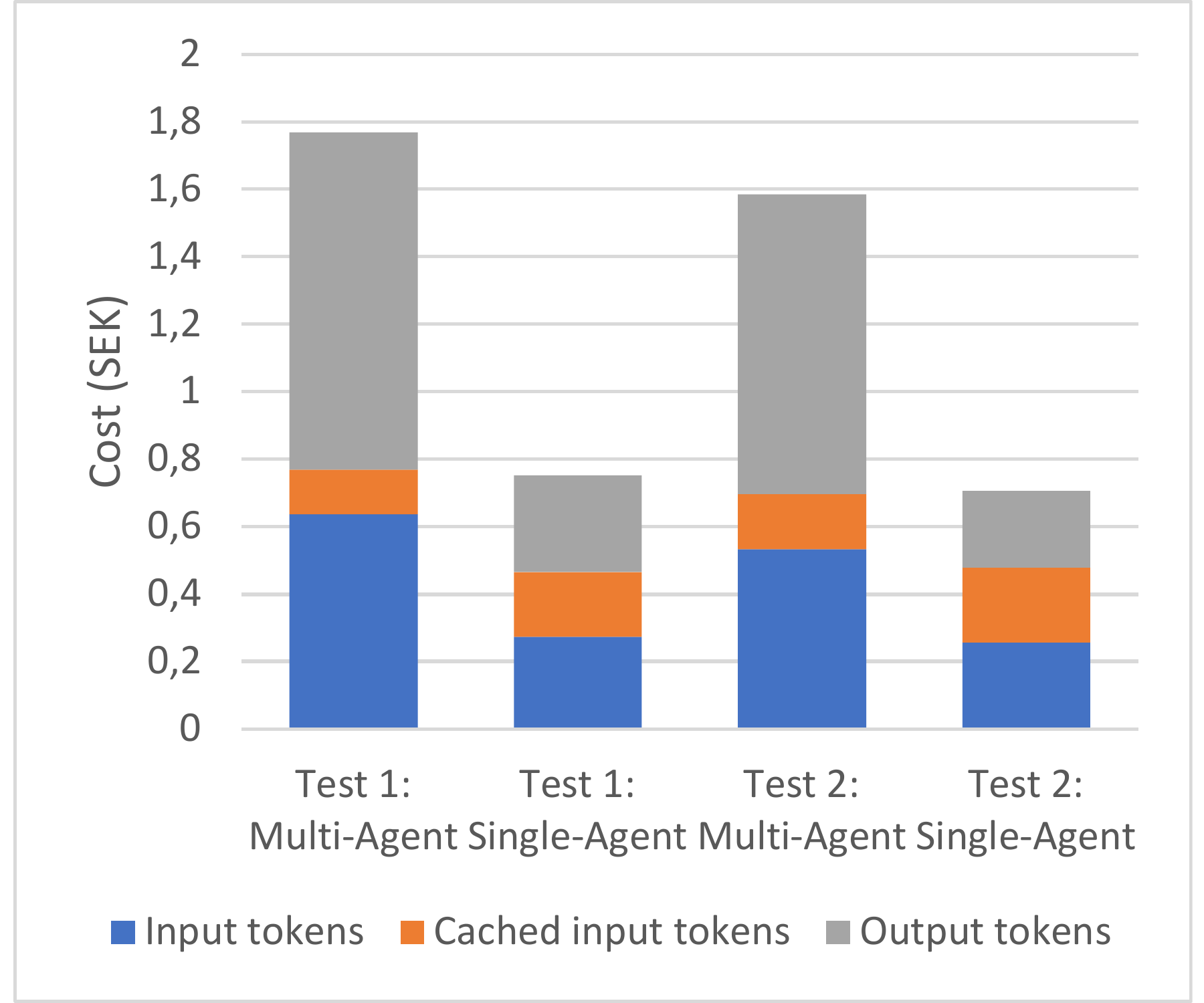}
        \caption{Cost contribution by token type}
        \label{fig:token_type_cost}
    \end{subfigure}
  \caption{Measurements gathered from 30 test analyses for each test-scenario and architecture combination. The X marks the mean; boxes show the percentiles, and the whiskers show minimum and maximum values.}
  \label{fig:dur_cost_results}
\end{figure}

\subsection{System-Level Metrics}
\label{sec:results_system_metrics}

System-level measurements were collected from 120 executions using the same scenarios as the practitioner survey. Each scenario--architecture combination was executed 30 times, resulting in 60 runs per architecture.
Figure~\ref{fig:consistency_results} shows that the single-agent system produced larger consistency groups in both scenarios, with the highest value observed in Scenario 2. 
Figure \ref{fig:dur_cost_results} shows that the single-agent required approximately 40 seconds per report, compared with about 130 seconds for the multi-agent. It also cost approximately 0.75 SEK (0.07 €) per run, while the multi-agent cost 1.5--1.75 SEK (0.15 €) and showed greater variation in execution time.
Although the single-agent used more total tokens, mainly cached input tokens, as shown in Figure \ref{fig:token_consumption}, Figure \ref{fig:token_type_cost} shows that uncached input and output tokens accounted for most of the higher multi-agent cost.

\subsection{Focus Group Results}
\label{sec:results_focus_group}

This section presents the qualitative findings from the focus group discussion. The findings are organized according to the predefined themes used in the focus group and analysis. These themes were used 
to structure the analysis of the transcribed discussion and to interpret how the participants understood, evaluated, and reasoned about the AI-generated RCA reports.

\subsubsection*{Theme 1: Understanding the AI's Reasoning.}
\label{sec:results_thematic_analysis_theme1}
\addcontentsline{toc}{subsubsection}{Theme 1: Understanding the AI’s Reasoning.}



Participants considered parts of the reports unnecessarily long and repetitive. The \textit{Root Cause} section sometimes restated error messages already presented as \textit{Evidence}, while references to tool behavior and retrieval limits were confusing and irrelevant to the failure itself. The \textit{Reasoning Steps} section was also perceived as unnatural:
\interviewquote{...I probably cannot explain my own reasoning steps, so reading it feels slightly unnatural.}

In contrast, the \textit{Next Steps} section helped participants understand how the agent interpreted the failure and what to investigate next.




\subsubsection*{Theme 2: Where the AI Provides Useful Guidance.}
\label{sec:results_thematic_analysis_theme2}
\addcontentsline{toc}{subsubsection}{Theme 2: Where the AI provides useful guidance}

Perceived usefulness depended on practitioners' experience and the complexity of the failure. Reports provided little value for familiar or simple failures.
However, participants considered them potentially more useful for less experienced practitioners or for unfamiliar failures. The \textit{Evidence} section was repeatedly identified as one of the most valuable parts because it directed attention to relevant logs before the proposed cause was accepted or rejected. Participants also suggested classifying whether a failure originated from the test framework or device software. The uncalibrated \textit{Confidence} value was difficult to interpret and participants were more likely to use the reports if they were integrated into the test framework and provided direct links to the referenced logs.

\subsubsection*{Theme 3: When the AI Leads Practitioners in the Wrong Direction.}
\label{sec:results_thematic_analysis_theme3}
\addcontentsline{toc}{subsubsection}{Theme 3: When the AI leads practitioners in the wrong direction}

Participants warned that plausible but incorrect conclusions could direct attention toward irrelevant causes and continue influencing the investigation after the report was set aside:

\interviewquote{...because it [the agent] says this looks normal, then you might not even think about looking there.}



Incorrect timestamps or time windows in the \textit{Evidence} section could also make referenced logs difficult to locate. Participants therefore expected that experience with the reports would be needed to understand their strengths and limitations.

\subsubsection*{Theme 4: Explanation Detail and Structure}
\label{sec:results_thematic_analysis_theme4}
\addcontentsline{toc}{subsubsection}{Theme 4: Explanation detail and structure.}

Participants generally preferred the most relevant information to appear first. The preferred entry point depended on the context: \textit{Root Cause} could be useful in meetings, while \textit{Evidence} was preferred when investigating failures in the test framework. Low or medium \textit{Confidence} values should also be visible early.

Less relevant sections, including \textit{Reasoning Steps}, \textit{Assumptions}, and \textit{Limitations}, could initially be hidden. Participants also recommended shorter root-cause descriptions, reduced timestamp precision, and time windows rather than isolated timestamps to improve readability.

\subsubsection*{Theme 5: Perceived Differences Between Architectures}
\label{sec:results_thematic_analysis_theme5}
\addcontentsline{toc}{subsubsection}{Theme 5: Perceived differences between architectures}

Participants generally perceived no systematic difference between reports from the two architectures:

\interviewquote{For me, you could have randomized this completely and I would not have seen any difference...}

A few participants preferred particular multi-agent reports, but this appeared to reflect their perceived correctness rather than an identifiable architectural characteristic. One multi-agent report also exposed an internal log retrieval limit that participants found confusing. This suggests that subagent behavior was not always fully hidden, although the study does not establish that this problem was unique to the multi-agent architecture.




\section{Discussion}

The results indicate two main findings. First, the evaluated architectures did not show a consistent difference in practitioner perceived report quality across the two scenarios. Second, the single-agent configuration provided a substantially better operational trade-off among execution time, cost, and semantic output stability. The following sections discuss these findings, their implications for tool design and industrial adoption and the main threats to validity.

\subsection{Architecture and Operational Trade-offs}
\label{sec:result_discussion_rq1}

The practitioner ratings were primarily scenario-dependent. The single-agent report received higher median ratings across all dimensions in Scenario~1, while the multi-agent report received higher ratings in Scenario~2. The practitioner evaluation therefore provides no consistent evidence that either configuration produced better reports. This interpretation is supported by the focus group in which participants generally reported few visible differences between the outputs.

The system-level measurements showed clearer differences. The single-agent configuration generated reports approximately three times faster and at about half the cost of the multi-agent configuration. It also produced a larger share of reports reaching the same underlying root-cause conclusion in both scenarios, as shown in Figure~\ref{fig:consistency_results}. This represents semantic output stability and not correctness, since repeated reports may consistently reach the same incorrect conclusion.

The difference in cost was not explained by total token usage alone. Figure~\ref{fig:token_consumption} shows that the single-agent configuration used more tokens overall, mainly because of cached input tokens. However, Figure~\ref{fig:token_type_cost} shows that the multi-agent configuration used more costly uncached input and output tokens. Its higher cost and longer execution time are consistent with the additional agent calls, intermediate outputs and coordination required between the orchestrator and subagents.

One possible explanation for the limited practitioner-perceived difference is that the evaluated scenarios and restricted tool set did not require the main potential strengths of a multi-agent design. Specialized agents may be more useful when an investigation involves a larger number of tools, substantial context or clearly separable responsibilities. In the present setting, the limited tool set reduced the need for specialization and coordination.

Context isolation also creates a trade-off. Subagents can process context-heavy tasks and return concise findings to the orchestrator, potentially reducing the amount of information handled in the main context. However, relevant details may be omitted or simplified during this transfer. The value of the multi-agent design therefore depends on whether subagent outputs preserve the necessary evidence without introducing high output-token cost or coordination overhead.

Neither configuration had access to the WeOS, test case or test framework source code and was therefore limited to diagnosing failures from logs and metadata. Access to source code, version-control history, runtime metrics, technical documentation or issue-tracking data could support deeper analysis, while also increasing the demands on tool selection and context management. The findings indicate that, for the specific implementations, tools and scenarios evaluated in this study, the single-agent configuration provided the better operational trade-off without a consistent disadvantage in practitioner-perceived report quality.

\subsection{Practitioner-Perceived Usefulness}
\label{sec:result_discussion_rq2}


The survey results show similar patterns for \textit{clarity} and \textit{trust}, as seen in Figure \ref{fig:T12SMA}. The \textit{usefulness} ratings varied more strongly and included the highest number of \textit{Strongly Disagree} responses. Focus group feedback indicates that perceived correctness influenced usefulness. However, the survey was not designed to investigate this association. In addition, the ratings may additionally reflect uncertainty about how the reports would fit into existing RCA workflows.

The focus group showed that usefulness depended on participants' experience 
and the quality of the generated conclusion. Experienced practitioners reported less need for assistance with familiar failures, while more complex or unfamiliar cases could increase the value of the reports.

Participants identified the \textit{Evidence} section as one of the most useful elements. Relevant log entries provided a starting point for further investigation, even when the proposed root cause was not accepted. This suggests that the system's practical value may lie more in evidence retrieval than in producing a definitive diagnosis.

Incorrect but plausible conclusions could also misdirect attention and increase investigation effort. Reports should therefore present conclusions as hypotheses supported by inspectable evidence, while practitioners retain responsibility for the final diagnosis.

\subsection{Ethical Considerations}

Because RCA reports may influence technical decisions, they should provide transparent and evidence-grounded explanations. Practitioners retain responsibility for accepting, questioning or rejecting the proposed conclusions. This is important because LLM outputs may appear plausible despite being incomplete or unsupported~\cite{deng2025deconstructing,high_level_overview_ai_ethics_kazim2021,strandberg2025ethical}.

The use of internal test data also requires appropriate access controls and agreements preventing submitted information from being stored or reused~\cite{deng2025deconstructing,ethics_of_ai_2022}. The practitioner evaluation followed informed consent, anonymization, restricted access, and recording deletion procedures~\cite{per_interview_ethics}. Future deployment should also consider the computational cost and energy use of repeated LLM calls~\cite{dangers_of_stochastic_parrots_2021}.

\subsection{Agent Implementation Differences}
\label{sec:discussion_agent_differences}

Both architectures used similar tools, context and instructions to improve comparability. However, this may have limited architecture specific optimization, since single and multi-agent systems may require different prompt and context designs. This particularly affected the multi-agent system, whose prompts were adapted from the single-agent baseline. Communication between the orchestrator and subagents may therefore not have been fully optimized. The multi-agent implementation should consequently be interpreted as a controlled variant rather than an optimized architecture.

During execution, the multi-agent system also occasionally triggered Azure's \textit{``jailbreak''} filter or produced invalid log query syntax. These issues may reflect overlapping instructions or incomplete context transfer between agents. Overall, the observations suggest that prompts, communication formats and tool instructions should be tailored to each architecture.

\subsection{Threats to validity}

Each participant rated one fixed report for each scenario--configuration combination.  Run-to-run variation and the procedure used to select the reports may therefore explain part of the observed results. The study also lacks an independently established ground truth and a timed manual RCA baseline. Therefore, correctness and usefulness represent practitioners' perceptions, while system duration reflects report-generation overhead rather than a demonstrated reduction in end-to-end RCA effort. The study does not show that either configuration is more accurate or faster than current manual practice. Participant familiarity with the test environment varied across roles, potentially influencing judgments of correctness and usefulness. Finally, our study compares one controlled implementation of each architecture rather than single-agent and multi-agent architectures in general. Cost and duration depend on the selected Azure deployment, model version, prompt-caching behavior and pricing during the experiment. The semantic-stability groupings also relied on an LLM evaluator without independent human validation.

\section{Conclusions and Lessons Learned}

This study evaluated LLM-based agents for supporting RCA of nightly test failures at Westermo. Neither architecture showed a consistent advantage in practitioner-perceived usefulness or report quality. Participants valued the reports mainly for locating relevant evidence and guiding further investigation.

The following lessons summarize the main implications derived from the practitioner feedback and system-level measurements.
\begin{mdframed}[
    linewidth=1pt,
    linecolor=black,
    backgroundcolor=white!10,
    roundcorner=5pt,
    innertopmargin=5pt,
    innerbottommargin=5pt,
    innerleftmargin=6pt,
    innerrightmargin=6pt
]
{\footnotesize
\textbf{Lessons Learned:}
\vspace{0.2cm}

\noindent
\textit{Start with a single agent baseline.}
The multi-agent configuration provided no quality advantage but was approximately three times slower and twice as costly.

\noindent
\textit{Prioritize evidence over definitive diagnoses.}
Participants valued relevant log evidence even when they rejected the proposed root cause.

\noindent
\textit{Design reports for the practitioner workflow.}
Detailed reasoning, repeated error descriptions, and internal tool information were considered unnecessary or confusing.

\noindent
\textit{Communicate uncertainty carefully.}
The confidence values were difficult to interpret, and plausible
but incorrect conclusions could misdirect investigations.
}
\end{mdframed}

Together, these lessons suggest that the value of LLM-based RCA lies primarily in supporting evidence discovery and further investigation. In the evaluated setting, a simple, evidence-focused single-agent configuration therefore represents the most appropriate starting point. These lessons remain specific to the studied environment, scenarios, participants, and implementations and should be evaluated further in broader and more complex industrial settings.

Future work should also investigate RCA workflows that prioritize evidence discovery and log navigation. This includes evaluating broader agent capabilities through source-code inspection and test analysis. Further research should also examine how structured RCA outputs can support other tasks, such as pull-request creation and automated bug fixing.

\section*{Acknowledgments}
This work is supported by the Swedish Agency for Innovation (Vinnova) through the project FLEXATION and by the Eureka Cluster on Software Innovation (ITEA) through the project MONA LISA.

\bibliographystyle{splncs04}
\bibliography{references}

\end{document}